\UseRawInputEncoding
\documentclass[12pt, letterpaper, twoside]{article}
\usepackage{graphicx}
\usepackage{subcaption}
\usepackage{fixltx2e}
\usepackage{hyperref}
\usepackage{multicol}
\usepackage{geometry}
\usepackage{nonfloat}
\usepackage{multicol}
\usepackage{float}
\usepackage{blindtext}
\usepackage{dblfloatfix} 
\usepackage{amsmath}
\usepackage{color}
\usepackage{epstopdf}
\usepackage{dsfont}

\makeatletter
\def\@seccntformat#1{%
\expandafter\ifx\csname c@#1\endcsname\c@section\else
\csname the#1\endcsname\quad
\fi}
\makeatother

\usepackage{setspace}
\definecolor{Sion}{rgb}{.45,0.05,.85}

\definecolor{green}{rgb}{.58,0.90,.20}

\usepackage{authblk}

\title{The growing topology of the \emph{C. elegans} connectome}

\author[1]{Alec Helm}
\author[1]{Ann S. Blevins}
\author[1,2,3,4,5,6,7]{Dani S. Bassett}
\affil[1]{Department of Bioengineering, School of Engineering \& Applied Science, University of Pennsylvania, Philadelphia, PA 19104 USA}
\affil[2]{Department of Electrical \& Systems Engineering, School of Engineering \& Applied Science, University of Pennsylvania, Philadelphia, PA 19104 USA}
\affil[3]{Department of Physics \& Astronomy, College of Arts \& Sciences, University of Pennsylvania, Philadelphia, PA 19104 USA}
\affil[4]{Department of Neurology, Perelman School of Medicine, University of Pennsylvania, Philadelphia, PA 19104 USA}
\affil[5]{Department of Psychiatry, Perelman School of Medicine, University of Pennsylvania, Philadelphia, PA 19104 USA}
\affil[6]{Santa Fe Institute, Santa Fe, NM 87501 USA}
\affil[7]{To whom correspondence should be addressed: dsb@seas.upenn.edu}
\date{}     

\begin{document}

\maketitle

\newpage
\begin{abstract}
Probing the developing neural circuitry in \textit{Caenorhabditis elegans} has enhanced our understanding of nervous systems. The \textit{C. elegans} connectome, like those of other species, is characterized by a rich club of densely connected neurons embedded within a small-world architecture. This organization of neuronal connections, captured by quantitative network statistics, provides insight into the system's capacity to perform integrative computations. Yet these network measures are limited in their ability to detect weakly connected motifs, such as topological cavities, that may support the system's capacity to perform segregated computations. We address this limitation by using persistent homology to track the evolution of topological cavities in the growing \textit{C. elegans} connectome throughout neural development, and assess the degree to which the growing connectome's topology is resistant to biological noise, such as variation in developmental timing. We show that the developing connectome topology is both robust to changes in neuron birth times and not captured by standard growth models. Additionally, we quantify the consequence of a neuron's specific birth time and ask if this metric tracks other biological properties of neurons. Our results suggest that the connectome's growing topology is a robust feature of the developing connectome that is distinct from other network properties, and that the growing topology is particularly insensitive to the exact birth times of motor neurons. By utilizing novel measurements that track biological features, we anticipate that our study will be helpful in the construction of increasingly accurate models of neuronal development in \textit{C. elegans}.
\end{abstract}

\newpage
\section{Author Summary}
Network analyses have identified several local and global properties of the \textit{C. elegans} connectome that are relevant to the organism's function and its capacity for information processing. Recent work has extended those investigations by focusing on the connectome's growth, in an effort to uncover potential drivers of connectome formation. Here we investigate connectome growth from the perspective of applied algebraic topology, by tracking both transient and persistent homology. In doing so, we are able to measure the resilience of the growth process to perturbations, and assess spatial variations in that resilience throughout the organism's body. We have herein determined the existence of a robust and topologically simple network feature that is unexplained by the presence of other notable features of the connectome. Our findings provide new insights regarding the development of topology in this simple natural connectome.

\newpage

\section{Introduction}

Model organisms such as the nematode \textit{Caenorhabditis elegans} proffer opportunities to understand brain development (and degeneration \cite{cadlwell2020modeling}) in a manner and at a scale that is inaccessible in the human \cite{rapti2020perspective}. The recent mapping \cite{varshney2011structural} of connections and neuron birth across time in \textit{C. elegans} provides a unique glimpse into the development process. Researchers can quantify patterns within the organism's brain development by using network science to encode the system as a network of nodes (representing neurons) and edges (representing connections) \cite{bullmore2009complex}. Studies combining network science and neuron-level developmental data have revealed the system's small-world characteristics \cite{amaral2000classes, latora2003economic}, the existence of a rich club of densely connected neurons \cite{towlson2013rich,arnat2018hub,colizza2006detecting}, and a sub-optimal wiring cost as defined by the length of connections between neurons \cite{perez2009structure}.

Most statistical quantities in network science, such as the degree distribution, modularity, and clustering coefficient \cite{costa2007characterization}, are naturally tuned to describe the structure of existing connections. Yet, present edges alone do not capture the richness of the network's topology. Instead, and for a more comprehensive view, it is of interest to also study absent edges. The \emph{absence} of connections between multiple neurons can be particularly important for many informational processes, such as isolating functions or providing parallel routing. A sparse area within a network might manifest within a pattern of four or more neurons connected as a loop, which can induce periodic activity patterns \cite{brunel2003determines, popovych2011delay} and support memory \cite{foss1996multistability,y1952histologie,ju2018network}. Precisely how the pattern of sparse areas might emerge over development could be susceptible to biological noise, as variations in neuronal birth times would induce variation in the order in which such voids emerge in the network. Further, variable birth times could lead not just to alterations in when voids occur, but also to whether they occur at all. For the purpose of this paper we refer to the variation of neural birth times as biological noise, and acknowledge that other sources of noise would be of interest to consider in future studies. 

Here we will investigate any absences or voids that appear within the growing neuronal network of \textit{C. elegans} and then quantify the extent to which noise alters these features. In considering which mathematical approach to use for our study, we note that the magnitude by which many network statistics could change does not scale with the magnitude of a perturbation induced by noise. This fact makes the study of noisy systems mathematically difficult. Our examination overcomes this limitation by capitalizing on mathematically precise methods to evaluate voids and their sensitivity to perturbation. In order to detect voids within the growing network, we use persistent homology \cite{zomorodian2005computing,carlsson2009topology,edelsbrunner2008persistent}, which records the voids, also known as cavities, that emerge, persist, and collapse throughout the course of the growing connectome's development. Importantly, the amount by which the persistent homology of a growing graph could change after perturbation has a strict upper bound that depends on the perturbation magnitude \cite{cohen2007stability}. Therefore, chronicling the evolving topology with persistent homology provides a method by which we can both (i) detect the development of sparse regions in the network that manifest as topological voids and (ii) quantitatively evaluate the robustness of these topological features to noise \cite{blevins2019reorderability,fasy2014confidence}.

In this paper, we test the hypothesis that the developing \textit{C. elegans} connectome produces evolving cavities, and that the evolving topology is robust to perturbations in neuronal birth times. We evaluate this hypothesis against an alternative hypothesis that the evolving topology is sensitive to perturbations, similar to the way a computation might be sensitive to the order of neuron activation. To do so, we analyze the persistent homology of the \textit{C. elegans} connectome as it develops from the first neuron born through to the complete adult network. We further analyze the stability of the persistent homology to perturbation of neural birth time and ordering, and find that the homological signature of \textit{C. elegans} changes little after the addition of noise. Further, we find that the persistent homology signature of the \textit{C. elegans} connectome is not reproduced by several conventional growth models, underscoring the need for future work to develop alternative models that better fit the data. Taken together, our evidence suggests that the existent persistent topology is an important and novel feature of the developing connectome in the nematode.

\section{Results}
Using data from \cite{varshney2011structural}, we construct a binary, undirected network that encodes the neural system. Each node in the network represents a neuron in the hermaphrodite \textit{C. elegans} and each edge represents the presence of a synapse or gap junction between the corresponding neurons. For recent reviews describing the biology of these two types of connections in \textit{C. elegans}, see Refs. \cite{jin2020gap,jin2015unraveling}. The full network consists of 279 nodes and 2287 edges. We expand this network into a growing graph by incorporating information about each neuron's birth time \cite{varier2011neural}. Specifically, we add each node to the growing graph at the time of the corresponding neuron's birth, and we add each edge as soon as both parent nodes are born. This process results in a many-stage growing graph, beginning with a single neuron and ending with the full adult connectome.

\subsection*{Cavities in the Growing Connectome}
\begin{figure*}[ht]
\centering
\includegraphics[width= 375 px]{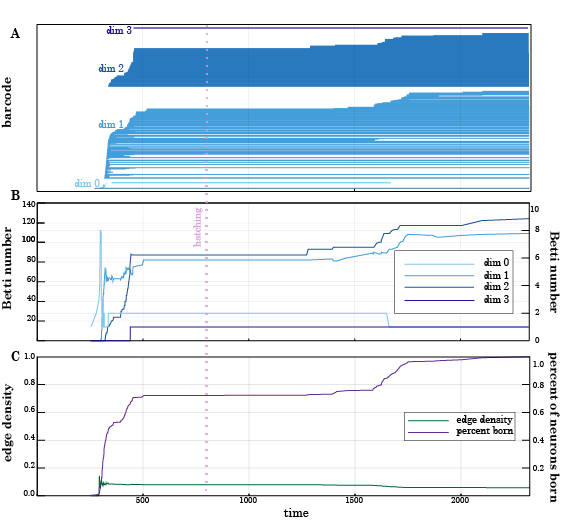}
\caption{\textbf{Many persistent cavities emerge in the developing neuronal connectome of \textit{C. elegans}.} \textbf{(A)} The barcode of the growing connectome. Each line corresponds to a persistent cavity in the growing connectome, colored by the dimension (see panel (B)). \textbf{(B)} Betti curves for dimension zero, one, two, and three. The Betti curves for dimensions one and two are marked using the left y-axis, whereas the Betti curves for dimensions zero and three are marked using the right y-axis. At each point in time, $t$, the height of the curve for dimension $n$, is equal to the number of cavities of dimension $n$ present in the connectome at time $t$. \textbf{(C)} Edge density (green) and the percentage of neurons born (purple) throughout the growth of the connectome. At time \textit{t = 0}, zero percent of the neurons are born; by time \textit{t = 2324.7}, 100\% of neurons are born. The vertical dashed line indicates the time at which the organism hatches. An explanation for the definition of cavities and barcodes is available in the Methods section (Fig.~\ref{fig:ph_introduction}).}
\label{fig:persistent_homology}
\end{figure*}

Before investigating the robustness of the developing connectome to changes in neuron birth order, we first seek to understand the evolving topology of the unperturbed growth process. As described in the Methods section, we compute the persistent homology of the growing connectome in dimensions zero through three. Note that a persistent cavity in dimension 0 corresponds to a connected component; a persistent cavity in dimension 1 corresponds to an un-tessellated loop between four or more neurons; and each higher dimensional cavity corresponds to an empty shell or capsule in the growing connectome (Fig.~\ref{fig:cliques_and_cavities}). In Fig.~\ref{fig:persistent_homology}, we show the growing connectome's barcode and Betti curves, two common statistical summaries of the persistent homology. Each line in the barcode (Fig.~\ref{fig:persistent_homology}A) corresponds to a persistent cavity in the growing connectome, with the bar extending along the time period in which the cavity persists. The Betti curve $\beta_n$ is a summary of the barcode of dimension $n$; the value of $\beta_n$ at any point in time $t$ is the number of topological cavities of dimension $n$ present at that point in time (Fig.~\ref{fig:persistent_homology}B). 

From the barcodes and Betti curves, we notice that early in development there are several disparate connected components. Fairly quickly, the majority of those components coalesce into a single component (Fig.~\ref{fig:persistent_homology}B). At 335.19 minutes a second component forms and persists for over a thousand minutes, but this isolated component only ever consists of six neurons. As expected, the connectome is nearly a connected graph for most of development, with slight deviations due to a small set of neurons (Fig.~\ref{fig:persistent_homology}C). Turning to higher dimensions, we see that the number of both first and second dimensional cavities tends to increase with time (Fig.~\ref{fig:persistent_homology}B). We observe that the neural development of \textit{C. elegans} involves the formation of many first dimensional cavities, and such cavities are both formed and destroyed at a high rate (see the barcode in Fig.~\ref{fig:persistent_homology}A). The Betti curve shows that construction and maintenance of cavities tend to prevail over the rate of death of cavities during early development (Fig.~\ref{fig:persistent_homology}A).

The birth of second dimensional cavities begins after that of many first dimensional cavities (Fig.~\ref{fig:persistent_homology}A). The embryonic rate of second dimensional cavity birth is quite high, and continually increases until suddenly stopping for nearly a thousand minutes. Quite unlike first dimensional cavities, we see from the barcode that almost all second dimensional cavities persist indefinitely (Fig.~\ref{fig:persistent_homology}A). Notably, despite the higher number of persistent cavities in dimension two than in dimension one for most of development, the barcode shows that the birth and death rates of persistent cavities in these dimensions differ dramatically. Second dimensional cavities emerge at a slower rate than do first dimensional cavities, but far fewer die over the course of neural development. Interestingly, third dimensional cavities do not appear until nearly three quarters of the neurons are born. At that time, a single cavity emerges and persists indefinitely. Due to the marked absence of third dimensional cavities, and the computational cost of finding these cavities, we do not perform further investigations into persistent homology beyond dimension two. 

Taken together, our results show that the \textit{C. elegans} developing connectome creates many persistent cavities during growth. We observe the highest rates of cavity formation and destruction during embryonic (pre-hatching) development, with first dimensional cavity activity starting to stabilize as second dimensional activity spikes. Post-hatching, there is a slow trend of cavity formation with the destruction of very few cavities that formed before hatching.

\subsection*{Topological Resilience to Reordering}

Noise affects processes at various biological scales \cite{davis2015low,skoczenski1998neural,cabral2017functional}, yet healthy systems overcome this unpredictable variation and proceed unscathed \cite{rudrauf2014structure}. In the context of our study, we expect this resilience to manifest in global structural properties of the growing connectome being largely unaffected by small, local perturbations. Specifically, we will ask how small changes in neuron birth time affect the persistent homology of the developing connectome. Further, we will evaluate the relationship between the temporal magnitude of the birth time change and the magnitude of the resulting topological change.

\subsubsection*{Birth-Time Noise}
\begin{figure*}[h]
\centering
\includegraphics[width= 0.8\textwidth]{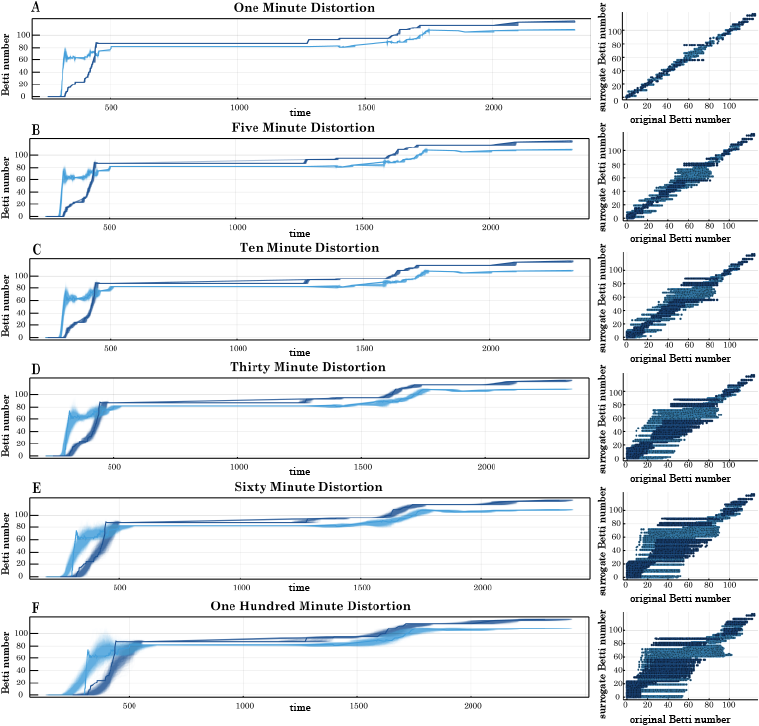}
\caption{\textbf{Betti curves remain largely stable with the introduction of birth-time noise.} Betti curve outputs of surrogate growing connectomes in which noise was added to neuron birth times (left). The true dimension one and two Betti curves are displayed in bold, and each of one-thousand surrogate curves are displayed in a transparent line above the original curve. Surrogate connectomes formed by re-assigning the birth time of each neuron to a new value no more than one minute \textbf{(A)}, five minutes \textbf{(B)}, ten minutes \textbf{(C)}, thirty minutes \textbf{(D)}, sixty minutes \textbf{(E)}, or one hundred minutes \textbf{(F)} from its original birth time. For each panel, we show a scatter plot (right) of Betti number pairs taken from their corresponding graphs. In each dimension (1, 2), for every surrogate curve and every minute (starting at the first minute a neuron is born and ending at the last minute a neuron is born), there is a point; the x-coordinate of that point is the true Betti number and the y-coordinate of that point is the Betti number of the surrogate curve (in the indicated dimension). Results of statistical analyses of these relationships are included in the supplement (Fig.~S3).}
\label{fig:noisy_birth_times}
\end{figure*}

In the simplest scenario, we introduce noise to the system by allowing the birth times of neurons to vary from their reported times. We then ask whether the persistent homology of the connectome changes in response to this type of perturbation. By altering the time that neurons are born, it is possible for cavities to form that do not appear in the course of the original connectome growth; it is also possible for some cavities that actually appear during development to instead never appear in the perturbed development. Since formation and destruction of topological cavities is determined by the birth time for our model, we employ a simple test of the stability of the Betti curve with respect to the addition of noise to neuron birth time. For each neuron we randomly assign a new birth time within a window of $\pm t'$ minutes around its reported birth time, compute the persistent homology, and repeat for 1000 surrogates. In order to get a broad understanding of how the amount of noise would affect the system, we report results for $t' = 1, ~5, ~10, ~30, ~60,$ and $100$ minutes (Fig.~\ref{fig:noisy_birth_times}). 

We observe that birth time can be quite noisy without qualitatively disturbing the Betti curve of the growing connectome (Fig.~\ref{fig:noisy_birth_times}). Up until thirty minutes of distortion, the surrogate curves tend to strongly replicate the general pattern of the true Betti curve. For the sixty and one-hundred minute cases there is more noticeable variation, as the surrogate curves tend to fall around a smoother reduction of the true curve. In order to quantify the differences between the true Betti curve and the collection of surrogates, we perform two tests of fit for the surrogate Betti curves to the true curve. As a first test, we calculated the average distance from the true Betti curve to each surrogate's curve at any point in time (Fig.~\ref{fig:noise_birth_times_stats}). Significance tests for fit of the true Betti curve against each sample of surrogate curves revealed that while in every case the addition of noise does introduce a statistically significant change to the Betti curve, in most instances this difference is exceptionally small. We then examined the correlation between the true Betti curve and each surrogate curve across development. At each minute, starting with the earliest birth time of neurons across surrogates and ending with the last such birth time, we recorded the Betti number in dimensions one and two of the true curve and each surrogate curve. Then for each minute and each surrogate we paired the true Betti numbers to the surrogate's Betti numbers. The Spearman correlations between the true and surrogate Betti number values were calculated (Fig.~\ref{fig:noisy_birth_times}, Fig.~\ref{fig:noise_birth_times_stats}). In all cases we found that $\rho > 0.90$ and $p < 0.0001$, indicating a significant correlation between the distorted values and the true values for Betti numbers across time. From these data, we conclude that Betti curves remain largely robust to small perturbations in the birth time of neurons.

\subsubsection*{Node-Order Noise}
\begin{figure*}[ht]
\centering
\includegraphics[width=\linewidth]{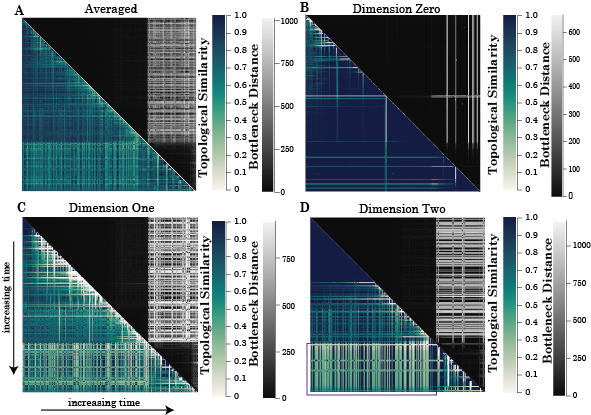}
\caption{\textbf{Assessing topological reorderability of the \textit{C. elegans} connectome.} Bottleneck distances (gray colormap) and topological similarity (green colormap) for every pair of neurons $(i,j)$. We show values \textbf{(A)} averaged over dimensions, \textbf{(B)} in dimension 0, \textbf{(C)} in dimension 1, and \textbf{(D)} in dimension 2. For all panels, the main diagonal of each matrix consists of ($i$,$i$) pairs, which have a topological similarity of 1 and a bottleneck distance of 0. Generally, darker colors indicate topologically similar nodes, whereas lighter colors indicate topologically dissimilar nodes. We have further added a notation on panel (D) to indicate the section of so-called mixed-pair neurons.}
\label{fig:topological_similarity}
\end{figure*}

At the most fundamental level, noisy birth times could induce topological changes to the growing connectome by permuting the neuron birth order. Continuing with our expectation that this system is resilient to small amounts of noise, we hypothesize that the topology of the growing connectome will change less after a swap between temporally close nodes than after a swap between temporally distant nodes. In order to test our hypothesis, we compute the bottleneck distance between the original persistent homology and the persistent homology after the node swap. Swaps that markedly change the persistent homology will result in a large bottleneck distance (see Methods; Fig.~\ref{fig:topological_similarity_supplement}), whereas swaps that do not appreciably change the persistent homology will result in a small bottleneck distance. The temporal distance, or time between swapped nodes' birth times, defines the maximum amount that the persistent homology \textit{could} change \cite{cohen2007stability}. To account for this upper bound, we additionally compute the topological similarity \cite{blevins2019reorderability} of each node pair, which normalizes the bottleneck distance by the temporal distance, and subtracts that value from one (see Methods; Fig.~\ref{fig:topological_similarity_supplement}). The resulting topological similarity between two nodes will be near one if their swap changes the persistent homology very little relative to what is possible, and near zero if their swap alters the persistent homology near maximally. We show the calculated bottleneck distances and topological similarity values for each neuron pair in Fig.~\ref{fig:topological_similarity}; each $(i,j)$ entry shows the corresponding value for the $(i,j)$ swapped neuron pair.

We first ask whether topological similarity indeed decreases as temporal distance increases, as expected. To answer this question, we calculate the Pearson correlation coefficient between the temporal distance and the topological similarity. In dimensions zero and one, we do not observe any strong correlation between these quantities (Pearson $\rho= 0.0796$, $p<0.001$ for dimension zero; $\rho=-0.0377$, $p=0.090$ for dimension one, with $df = 38779$ in all cases). In dimension two, we see a weak negative correlation ($\rho = -0.0399$, $p<0.001$). These results demonstrate that the topological similarity is either not at all associated with temporal distance or is only weakly associated. Thus, our results do not support our original hypothesis of a simple negative relation between temporal distance and topological similarity.

Interestingly when we consider patterns of relationships, we observe that in all dimensions, the bottleneck distances (Fig.~\ref{fig:topological_similarity}) show a striking trend that is defined by the birth of the neuron PQR. Before describing this trend, we will first note that we refer to a pair of neurons where one is born before PQR and the other born after PLML as a \emph{mixed pair}, and we refer to other pairs of neurons as \emph{unmixed pairs}. We find that both types of unmixed pairs --- pairs of neurons where both are born before the $202^{nd}$ neuron (PQR) and pairs where both are born after the $201^{st}$ neuron (PLML) --- appeared to have lower bottleneck distances than mixed pairs. The section of topological similarity scores for mixed-pair neurons is noted in Fig.~\ref{fig:topological_similarity}D. We performed a Mann-Whitney $U$-test to determine whether a significant difference exists between the topological similarity of mixed and unmixed pairs of neurons, separately for each dimension for a total of 3 tests. We then repeated the three tests using the bottleneck distances instead of the topological similarities. In all six tests, the $U$ statistic was on the order of $10^{8}$ ($p < 0.001$), allowing us to reject the null hypothesis that no significant differences exist. It is possible that this difference between the mixed and unmixed neuron pairs obfuscates any more subtle relation between topological similarity and temporal distance. 

To further understand this result, we notice that the time gap between the birth of PLML and the birth of PQR is 770.25 minutes, notably longer than the 4.70 minutes that on average separates all other temporally adjacent neuron pairs. This large time interval could have a marked impact on our findings: it would be appended to any bottleneck distance between unmixed pairs in which the swap caused a persistent cavity to move across that time interval, thus driving up the bottleneck distance and driving down the topological similarity. To determine whether our results are influenced by the presence of this large time gap, we performed a time-gap normalization. Specifically, we calculated the topological similarity for a growing graph that is identical to the original, except that we reduce the birth time of all neurons born after PLML by 765.55 minutes. This change reduces the time gap between the birth of PLML and PQR to the average time gap observed between all other temporally adjacent neurons. 

After performing this time-gap normalization, we again calculated the Spearman correlations between the topological similarity and temporal distance. We found that the dimension zero and two coefficients remained quite similar ($\rho= 0.048$, $p<0.001$, and $\rho = -0.338$, $p<0.001$, respectively), but the first dimension case changed sign and strengthened ($\rho= 0.270$, $p<0.001$). The time gap normalization therefore does not reveal a strong relation between topological similarity and temporal distance, contrary to our initial hypothesis. Furthermore, the time gap normalization did not alter the significant difference in topological similarity between the mixed and unmixed neuron pairs. Specifically, we again used a Mann-Whitney $U$-test to evaluate the difference in topological similarity (or bottleneck distance) between the mixed and unmixed neurons for all dimensions. We found a significant difference between the groups ($p = 10^{-99}$ for all but the topological similarity in dimension zero where $p = 10^{-43}$, and in all cases the $U$ statistic was on the order of $10^{9}$). This pattern of findings indicates that the topological separation of the groups is not merely an artifact of the large time interval, but that neurons born across the hatching barrier exhibit some significant lack of topological reorderability.

\subsection*{Comparisons to common growing network null models}

To assess whether our observations are specific to \textit{C. elegans} or more generally expected in developing networks, we consider common growing network null models that capture different properties of the \textit{C. elegans} system. We use four models in this paper: Erd\H{o}s-R\'{e}nyi, spatial growth, configuration, and economical spatiotemporal growth (ESTG) \cite{nicosia2013phase}; for definitions of all four models, see Methods. The Erd\H{o}s-R\'{e}nyi model produces a random graph that approximately preserves the edge density of the true connectome; the spatial growth model produces a random graph that has identical edge density to the true connectome and minimizes wiring cost given random node placement; the configuration model produces a random graph with a degree distribution identical to the true graph; and the ESTG model produces a random graph that favors low wiring cost and formation of connections to high-degree nodes. The first three are standard null models that capture some of the important properties of the true connectome (edge density, wiring cost, and degree distribution), and the fourth model is known to produce a final graph that shares many of the properties found in the true connectome \cite{nicosia2013phase}. 

\begin{figure*}[ht]
\centering
\includegraphics[width=\linewidth]{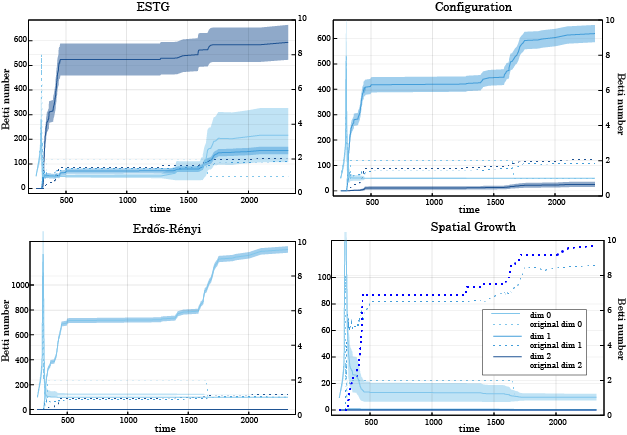}
\caption{\textbf{Common growing network null models largely fail to predict the persistent homology of the \textit{C. elegans} developing neural network.} For all four of our null models, we show the averaged Betti curves for dimension zero, one, and two across the 1000 surrogate curves. Shaded areas correspond to $\pm$ the standard deviation across trials. Betti curves for the true connectome are plotted with dashed lines.}
\label{fig:null_model_persistent_homology}
\end{figure*}

We first sought to determine whether the null models produce Betti curves similar to the reported connectome. To do so, we constructed 1000 surrogate connectomes for each null model and then computed their Betti curves in dimensions zero through two (Fig.~\ref{fig:null_model_persistent_homology}). It is immediately apparent that the Erd\H{o}s-R\'{e}nyi, spatial growth, and configuration models fail to replicate the dimension one and two persistent homology found in \textit{C. elegans}. Erd\H{o}s-R\'{e}nyi produces no dimension two homology and excessive dimension one homology. The configuration model does similarly, though to a lesser extent in both instances. The spatial growth model produces far less homology in both dimensions. All three fairly accurately capture the dimension zero homology; that is, the number of connected components. The ESTG model captures the true dimension one persistent homology almost perfectly, indicating that the growth model correctly predicts the existence of loops in the connectome. However, the ESTG model curiously produces approximately five times more second dimensional cavities than the true connectome. We also note that the ESTG model does not replicate the true zeroth dimensional homology. The ESTG model approximately captures this feature up until around 1700 minutes, but after that time the model starts producing new connected components that never rejoin the main component. On average the ESTG model produces a connectome made of three disjoint components, as opposed to the true connectome which is made of a single connected component. 

We further wished to determine whether the null models produce growing connectomes with topological resilience similar to that of \textit{C. elegans}. To do so, we produced one hundred surrogate connectomes for each null model. For each of these surrogate connectomes we performed birth-time distortions with time maximums of 1 minute, 5 minutes, 10 minutes, 30 minutes, and 60 minutes, as before, and found the distance between each distorted connectome and its surrogate. This process produced 100 area difference measures for each of the 100 surrogate connectomes. A two-sample $t$-test was performed to determine whether these distortion-induced differences were similar to the differences found in the true connectome. In all cases (across dimensions, distortion ranges, and null models) the $p$-value for the $t$-test was less than $0.001$ with a test statistic greater than $3.7$, indicating that the null models produce similarity distributions that are distinct from the true distribution. In all cases the test statistic decreased with increasing distortion, and was quite a bit larger in the first and second dimensional cases than in the zeroth.

We then wished to determine whether the null models produce connectomes where the individual neurons display similar trends of impacting the system's resilience to change. For each null model, we therefore constructed four surrogate connectomes for topological similarity calculations. We calculated the topological similarity matrix for each surrogate, and then found the difference between each of these matrices and the true topological similarity matrix. The average of the similarities for each trial, as well as the variance across the four trials, are displayed in Fig.~\ref{fig:null_model_topological_similarity} and Fig.~\ref{fig:null_model_top_sim_supp}. We then performed a one sample $t$-test on the set of differences between the topological similarity scores for each node pair in the null models and the true connectome to determine whether the null model's topological similarity scores were similar to those from the topological similarity of \textit{C. elegans}. In all cases the $p$-value for the $t$-test was less than $10^{-20}$, indicating that the topological similarity signature of these null models is quite distinct from that of the reported connectome. Written in order of increasing dimension, the $t$-statistics for the configuration trials were -22, 523, and 358; for ESTG the values were 10, 21, and -214; for Erd\H{o}s-R\'{e}nyi the values were 39, 52, and 385; and for spatial growth the values were -9, 516, and 358.

\begin{figure*}[ht]
\centering
\includegraphics[width=\linewidth]{Figure5.png}
\caption{\textbf{Persistent homology of common growing network null models.} For both the configuration and the ESTG null models, we show the average of the topological similarity across the four trials (in shades of green) and the standard deviation of the topological similarity values across the trials (in the map from yellow to purple). Similar images for the Erd\H{o}s-R\'{e}nyi and spatial growth models can be found in the Supplement (Fig.~\ref{fig:null_model_top_sim_supp}).}
\label{fig:null_model_topological_similarity}
\end{figure*}

We conclude this section by noting that the topological signatures we observed in the growing neuronal connectome of \textit{C. elegans} are not generally expected in common growing network models. Thus we may conclude that the topological signatures herein identified may represent organism-specific biological principles of growth, evolution, and function, as opposed to mere artefacts of conventionally important network features.

\subsection*{Relations to Biological Properties}

In a final investigation, we sought to understand how biological properties of individual neurons relate to the global topological signatures that we observe in the developing neuronal connectome of \emph{C. elegans}. The above analyses have been limited in that we treated each node as an internally identical piece of a growing connectome; yet in truth, neurons differ appreciably from one another in terms of morphology, connectivity, and function. We thus sought to determine whether biological qualities exhibited by different classes of neurons might relate to topological qualities exhibited by their corresponding nodes in the connectome. Specifically, we ask whether neuron type (motor, sensory, interneuron) or location (head, body, tail) are related to either the number of persistent cavities formed or killed by the corresponding node, or to the summed topological similarity of the node \cite{white1986structure}.

We begin with our first question. Is neuron type or location related to the number of persistent cavities formed or killed by the corresponding node? To answer this question we note that the birth of each neuron is a movement from one stage to the next in our growing connectome, and therefore each neuron has an associated set of topological cavities that it tessellates away and to which it gives rise. In order to determine whether a trend exists between biological features and topological cavity formation or tesselation, we constructed 1000 surrogate connectomes using imposed random linear ordering on birth times (see Supplementary Methods for details and motivation). We then counted the average number of cavities in dimensions zero through two to which each neuron's birth gives rise and terminates across the trials. We compared these average counts to neuron type (motor, sensory, or inter) and location (head, body, or tail). We note that for dimensions one and two, most neurons have no part in cavity formation or destruction. Indeed, only $42\%$ of neurons take part in dimension one cavity formation or destruction, and $20\%$ of neurons take part in dimension two cavity formation or destruction. In dimension one, we observe that a greater proportion of interneurons take part in cavity termination than the other neuron types. Additionally we find that head neurons tend to tessellate more loops and to form more second-dimensional cavities than neurons in other regions. These findings demonstrate that neurons of different types and bodily locations play distinct topological roles in the dynamic neuronal connectome of \emph{C. elegans}.

\begin{figure*}[ht]
\centering
\includegraphics[width=350px]{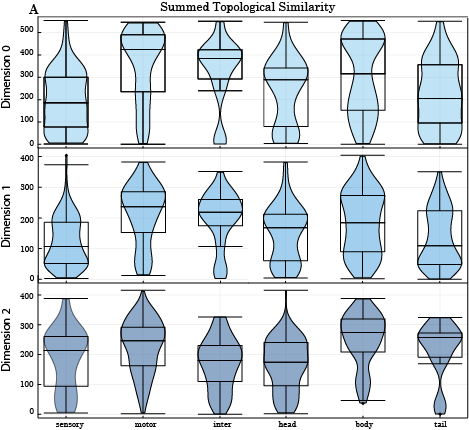}
\caption{\textbf{Summed topological similarity differs across neuron category types and bodily locations.} Boxplots and overlain violin plots for the distribution of summed similarity scores for each neuron sorted into bins both by neuron type (motor neuron, sensory neuron, and interneuron) and by region in the body (head, body, and tail). The summed similarity score for a neuron is the sum of the topological similarity that the neuron exhibits with all other neurons in the connectome. For statistical testing, we performed an analysis of variance (ANOVA), and we show full results of that test in Fig.~\ref{fig:linear_order_trials_supplement}.}
\label{fig:biology_top_sim}
\end{figure*}

We now turn to our second question. Is neuron type or location related to the summed topological similarity of the node? To answer this question, we assigned each neuron a similarity score equalling the summed topological similarity that it exhibited with all other neurons. Note that this measure is in direct proportion to a similarity score equalling the average similarity that the neuron exhibits with all other neurons. Intuitively, a neuron with a high summed topological similarity can be born at any of a host of times without having a drastic affect on the persistent homology. Conversely, a neuron with low summed topological similarity can greatly alter the topology if born at a time different from its true time. For each dimension, we separated neurons by type (sensory, motor, inter) or location (head, body, tail) (Fig.~\ref{fig:biology_top_sim}). Across all three dimensions, motor neurons had significantly higher summed similarity than other types of neurons. Interneurons tended to have lower summed similarity scores than other types of neurons across all three dimensions. To assess the statistical significance of these observations, we performed an analysis of variance (ANOVA) for each dimension and each grouping (type or location); for full tables, see Fig.~\ref{fig:linear_order_trials_supplement}. We found that the difference between any pair of types in any dimension is statistically significant, aside from the (interneurons, sensory neurons) pair whose difference was not significant in any dimension. Across all dimensions, body neurons tended to have higher than average summed similarity, and head neurons tended to have lower than average summed similarity. Differences by region categories were significant in every case aside from the differences between tail and body neurons (Fig.~\ref{fig:linear_order_trials_supplement}). 

In summary, our exploratory analysis revealed that neurons of different types or locations participate in the evolving topology in distinct ways. Both tests that we performed indicated that interneurons have a crucial topological role in the growing connectome, as they are involved in forming and terminating cavities, and are not readily swapped with other neurons.

\section*{Discussion}
Here, we examined the persistent homology of the \textit{C. elegans} connectome throughout development. We calculated the growing connectome's barcode and Betti curves, which showed that the connectome at first produces several connected components that eventually coalesce into a single component; that many untesselated loops form throughout development, a third of which do not persist in the adult worm; and that many second dimensional shells emerge, though at a slower rate, and almost always persist in the adult worm. We further observed that the trend of production and destruction of these cavities is robust to small perturbations to the birth times of individual neurons. We then showed that locally altering the order in which neurons are born tends to cause drastic changes in the underlying persistent homology, relative to the maximum possible change. The observed homological trends are not wholly explained by conventional growth models, although we found that the economical spatiotemporal growth model captures the first-dimensional homology relatively well. Finally, we showed that the amount by which birth time change affects the growing topology varies by neuron type and location, with motor neurons (sensory) tending to be the most (least) amenable to being born at the wrong time. Our work leaves open the question of how these persistent homology features might be encoded in the genetic program \cite{arnat2019uncovering}, but in focusing on the proximal biology of the circuitry our study points out several research directions that are immediately accessible to ongoing work.

\paragraph{Expanding the economical spatiotemporal growth model.}
We showed that the second dimensional persistent homology is a relatively robust feature of the growing connectome that is not wholly replicated by any of the growth models tested in this paper. In fact, the best performing model, the economical spatiotemporal growth (ESTG), matched the persistent homology in dimension one but not in dimensions two or zero. We suspect that the observed robustness of the dimension two persistent homology is driven by a latent feature of the biological system. Future work analyzing the ESTG model with respect to the generation of second dimensional cavities could provide insights regarding the growth logic of cavity generation in \textit{C. elegans}. Such work could also lay a foundation for determining the factors of growth models that have significant implications for resultant persistent homology. The distribution of topological similarity scores is a complicated metric of the node relations, and is likely perturbed in varying ways by the many controls utilized in growth models. Perhaps a large class of models exists that accurately reproduces persistent homology while respecting simple relative identifiers about the nodes, the latter of which is done by ESTG through factoring in neuron location, and thereby accurately mimics the topological similarity metric. Such relations between the constants in a growth model and outputs in persistent homology is an open field of inquiry upon which such research into the ESTG model may shed light.

\paragraph{Relation of topological voids to network dynamics.} Persistent homology, and topological data analysis more generally, has been recently used to study various networks and to answer pressing questions across a range of disciplines including but not limited to neuroscience \cite{salnikov2018simplicial,curto2017can,curto2008cell,kanari2020trees}. Studies have identified topological differences between distinct sets of neural connections such as those within brain areas that developed early in evolution versus those that developed late in evolution \cite{sizemore2018cliques}. The approach has also been used to identify topological differences in functional connections in the brain that are induced by drugs such as psilocybin \cite{petri2014homological}, as well as differences between functional brain networks of individuals with schizophrenia and those of their unaffected siblings \cite{stolz2018topological}. In parallel through this same mathematical lens, we have a better understanding of dynamical properties of networks of neurons \cite{morrison2016diversity}) as well as looped motifs within such systems \cite{campbell2004delayed}. Future work investigating the direct functional implications of various persistent homologies within neuroscience may more directly tie topology with experimental measurements such as calcium imaging and associated computational models of collective behavior therefrom \cite{chen2019searching,gerkin2018towards}. Such efforts could shed light on the complementary questions of (i) why there is a relatively high topological similarity across motor neurons, and (ii) why most neurons do not directly have a part in the formation or destruction of topological cavities.  

\paragraph{Methodological considerations and limitations.} Several methodological considerations are pertinent to this work. First, our analysis assumes that neuronal connections form as soon as all involved neurons are born. Although this assumption is untrue, it is useful as an approximation of development. It revealed that the reorderability of neuron births explains why topological structure is resistant to the stochasticity of continuous axon branching and guidance \cite{dent2003axon, razetti2018stochastic}. Future work could use data with greater spatiotemporal resolution to improve our estimates of the persistent homology. Second, we note that the robustness of the Betti curves to perturbations of neuron birth time indicate that delays in synaptogenesis would not drastically alter our results. However, it would be important to investigate this prediction directly in future work that incorporates experimental knowledge about time delays involved in functionally connecting neurons. Third, it would be useful in future work to extend our methods to focus on edge birth times as opposed to node birth times. Such an extension could incorporate connection formation times, and their associated delays, when such data becomes available. Fourth and finally, it would be of interest to extend these same methods to increasingly accessible data regarding sexual dimorphism in \emph{C. elegans} \cite{walsh2020what,mulcahy2018pipeline,oren2016sex}, developmental variation across other nematode species \cite{haag2018evolutionary}, and the development of growing connectomes in other animals, especially humans \cite{vertes2015annual,vertes2014generative}.

\paragraph{Future directions.} Our work motivates future efforts both in mathematical methods and in biology to model brain network development. In the context of the mathematical methods, our work represents the first biological test of the reorderability concepts introduced in Ref. \cite{blevins2019reorderability} (with inspiration from Refs. \cite{cohen2007stability,memoli2019quantitative}). Consequently, our study opens many new methodological questions whose answers would improve future analyses. First, how might we calculate the \emph{growing graph specific} upper bound for a change in persistent homology? Here we use a large number of randomized node birth times to provide a sense of the possible space of persistent homology outputs from a single graph. However, a defined upper bound would make comparing reorderability between two graphs significantly more rigorous. Additionally, identifying subgraphs in which the persistent homology cannot change could greatly reduce the computational time needed for pairwise swap reorderability calculations, as one could simply ignore node pairs that participate in such a completely reorderable subgraph. We look forward to future efforts in these directions.

In addition to methodological questions, our work opens new avenues for inquiry on the development of neurobiological networks. Using methods from Ref. \cite{yoon2020persistence}, one could ask how the topology evolves within connections among particular neuron types, and how this topology relates to or fits in with the evolving topology of the entire connectome. Furthermore, future work could further investigate the biological difference between the swappable (high summed topological similarity) and unswappable (low summed topological similarity) neurons. Although here we found that swappability differs by neuron type, we expect that this result does not carry the totality of the information about swappable nodes. Based on the clear white streaks in Fig.~\ref{fig:topological_similarity} and the subtle bimodal shape of the distributions in Fig.~\ref{fig:biology_top_sim}, we expect that a more optimal solution to separating the swappable and unswappable neurons exists than found in this work. Finally, our reorderability results provide new theoretical predictions about which neurons would most affect development when perturbed, which could be tested empirically with perturbation experiments \cite{waaijers2013crispr,paix2017precision}. Along these lines, it would be of interest to determine whether the neurons that most affect development when perturbed are also the neurons predicted to have greatest influence on function according to synthetic ablation studies \cite{towlson2020synthetic}.

\section{Conclusion}
In this work we investigated the evolution of evolving voids within the developing \textit{C. elegans} neuronal network. We found that many such voids exist, that these are stable features with respect to noise, and that neurons vary in their contributions to these voids across type and location. Together our results suggest that topological cavities are a robust feature of the developing \textit{C. elegans} connectome, and lay the groundwork for future biological and mathematical experiments that investigate the role of cavities within the growing brain.

\section{Methods}

\subsection*{Data and software}
We used data from Ref. \cite{arnat2018hub} to construct the \textit{C. elegans} connectome and assign neuron positions. We also used the data within celegans279dev.zip from the Dynamics Connectome Lab \cite{varier2010neural} to assign neuron birth times. We assigned neurons to the types ``sensory", ``motor",  or ``interneuron" based on Ref. \cite{hobert2016revisiting}, and we used data for length approximations during development from Refs. \cite{mckeown1998sma, altun2002wormatlas}. All scripts used for calculations and figure generation are publicly available on GitHub (https://github.com/AlecHelm/C-Elegans-Persistent-Homology) and rely on the Eirene package in Julia \cite{henselmanghrist16} and the TDA package in R \cite{fasy2014introduction}.

\subsection*{Graphical Representation}
\begin{figure*}[]
\centering
\includegraphics[width=\linewidth]{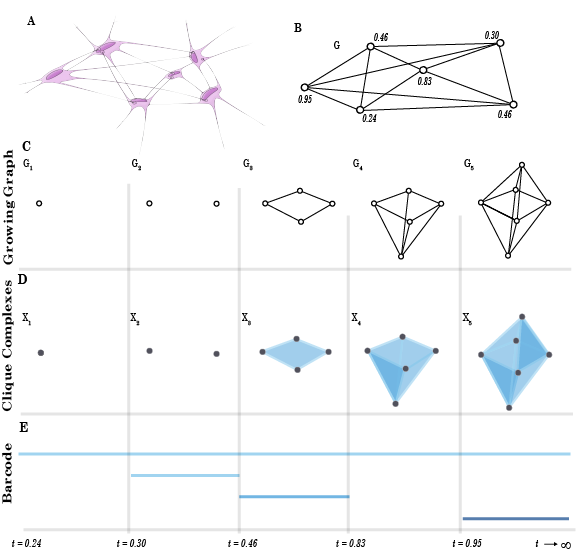}
\caption{\textbf{Tracking persistent cavities in a growing connectome.} \textbf{(A)} An example connectome of six neurons with sixteen synapses or gap junctions. Some neuron pairs exhibit multiple connections between them. \textbf{(B)} A graph reduction of this biological structure, where each neuron is a node and whenever connections exist between two neurons there is a single edge connecting their corresponding nodes. Furthermore, each node is labeled with a number representing the neuron's birth time. \textbf{(C)} A growing graph composed of five graphs. Each graph represents a stage in the growing graph, where we have added the next set of nodes by birth time, and any connections that exist in $G$ between nodes present at that stage. \textbf{(D)} The clique complex of each graph in panel (C). Each subset of all-to-all connected nodes is identified as a unit. Specifically, cliques between $k+1$ nodes are mapped to $k$-simplices. \textbf{(E)} The barcode for the growing graph. We have a single bar of dimension 0 present throughout, and a smaller bar of dimension 0 present only for the brief period where there is a second connected component. There are also bars of dimension 1 and 2 present when cavities of those dimensions are present.}
\label{fig:ph_introduction}
\end{figure*}

We begin with the construction of a series of graphs that will represent the growth of neural connectivity in the hermaphrodite \textit{C. elegans} as neurons are born (Fig.~\ref{fig:ph_introduction}). The raw data describes a connectome of 279 neurons and the 7283 connections that exist between neuron pairs in the adult \textit{C. elegans} neural system \cite{varshney2011structural}. Note that these 7283 connections include multiple connections between the same two neurons; we reduce these data to 2287 connections by treating multiple connections between the same two neurons as a single connection. A possible subset of this biological network is depicted in Fig.~\ref{fig:ph_introduction}A. We reduce this connectome to a graph in which nodes represent neurons and edges represent connections between neuron pairs in the adult nematode (Fig.~\ref{fig:ph_introduction}B). By including the neuron birth information from Ref. \cite{varier2010neural}, we construct a node-weighted graph $G = (V,E,b)$ where $V$ is the set of 279 nodes, $E$ is the set of 2287 edges, and $b$ is the function that assigns vertices to their corresponding neuronal birth times. More explicitly, we use the information from Ref. \cite{varier2010neural} to create an assignment $b: V \rightarrow \mathbf{R}$ in which $b(v)$ equals the birth time of the neuron corresponding to node $v$. 

Next, we mimic the growing process by constructing a telescoping series of sub-graphs of $G$ via the following inductive method. We begin by locating the earliest birth time $b_1$ assigned to any node (in our case, $b_1 = 253.08$ minutes). We construct $G_1$ by including from $G$ all nodes $v$ satisfying $b(v) = b_1$ and any edges in $G$ connecting these nodes. In our case, $G_1$ consists of a single node that represents the first neuron born (Fig.~\ref{fig:ph_introduction}). Given any graph $G_n$ that is not equal to $G$, we construct $G_{n+1}$ as follows. We locate those nodes in $G \setminus G_n$ with the earliest birth time, and we call this $b_{n+1}$. Then, $G_{n+1}$ is the graph consisting of all nodes with birth time $b_{n+1}$, all nodes in $G_n$, and all the connections that exist between these nodes in $G$. In this way, we generate a series of 172 sub-graphs $G_1 \subset G_2 \subset \dots \subset G_{172} = G$ where the first graph contains only the first neuron born, and each consecutive graph adds the next neuron(s) born and all connections that this neuron(s) makes with previously added neurons (and also all connections between the multiple neurons added within a step). The final graph in the sequence is identical to $G$. The construction of such a series of graphs is illustrated in Fig.~\ref{fig:ph_introduction}C.

\subsection*{Persistent Homology}
Our goal is to understand the evolving structure of the \textit{C. elegans} neuronal network in a mathematically rigorous way that still provides qualitative intuition. Persistent homology \cite{edelsbrunner1995union,zomorodian2005computing}, a method from applied topology, offers both a deep mathematical framework and interpretable insight regarding a system's structure and function. Specifically, persistent homology characterizes the evolution of topological features, called cavities or cycles, within the growing network. Network models display particular patterns, sizes, and rates of cavity evolution \cite{sizemore2016classification,horak2009persistent} and in neural systems, a topological cavity might indicate robustness of communication pipelines, parallel processing, or otherwise intentional sparseness within the network. For a general and more rigorous treatment of persistent homology and algebraic topology in general, we direct the interested reader to Refs. \cite{ghrist2008barcodes, cerri2013betti, otter2017roadmap}. We also include a more detailed and data-specific description of persistent homology in the Supplementary Methods.

In brief, in order to compute the persistent homology of the connectome we first translate our sequence of growing graphs $G_1 \subset G_2 \subset \dots \subset G_{172}$ into a sequence of \textit{clique complexes} $X_1 \subset X_2 \subset \dots \subset X_{172}$ (Fig.~\ref{fig:ph_introduction}D). A clique complex $X_i$ is a particular type of representation of the data in which any group of $k+1$ nodes that are all-to-all connected in $G_i$ (a $(k+1)$-clique) is a related unit, called a $k$-simplex. A 0-simplex is a node; a 1-simplex is an edge between a pair of distinct nodes; a 2-simplex is a set of three nodes with edges forming a triangle; and so on. From the viewpoint of algebraic topology, a simplex of any size is homologically identical, and the existence of non-trivial homology groups can only emerge from connections between distinct simplices in the larger clique complex.

\begin{figure*}[]
\centering
\includegraphics[width=\linewidth]{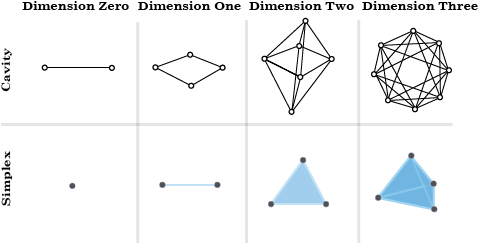}
\caption{\textbf{Examples of simplices and cavities.} The top row shows cavities in various dimensions. A cavity in dimension 0 corresponds to a connected component; a cavity in dimension 1 corresponds to an untessellated loop between four or more neurons; and each higher dimensional cavity corresponds to an empty shell or capsule in the growing connectome. The colored regions in the bottom row show k-simplices of size zero through three, each of which is an all-to-all connected set of nodes.}
\label{fig:cliques_and_cavities}
\end{figure*}

A \textit{$k$-cycle} is a loop composed of $k$-simplices (Fig~\ref{fig:cliques_and_cavities}). In order to detect the topological voids within the clique complex $X_i$, we mathematically compress out those $k$-cycles that do not surround any unique cavities. This operation defines an equivalence class of $k$-cycles for each cavity, such that all cycles surrounding the same cavity are considered equivalent. Distinct equivalence classes\footnote{specifically, those that form a basis for the space of equivalence classes} of $k$-cycles are in one-to-one correspondence with the topological cavities of dimension $k$. We then define the \textit{$k^{th}$ Betti number} to be the number of such equivalence classes, or equivalently the number of cavities in dimension $k$. 

Whereas homology can be used to find the topological cavities for a single clique complex, we use persistent homology to compute the beginning, evolution, and collapse of topological cavities as the system grows. More specifically, persistent homology allows us to track the persistence of cavities from $X_{i}$ to $X_{i+1}$, such that when we analyze our entire sequence of clique complexes $X_1 \subset X_2 \subset \dots \subset X_{172}$ that describe the growing neural system, we can produce a list of topological cavities that begin, evolve, and disappear as the \textit{C. elegans} neuronal network grows. For each persistent cavity, the persistent homology computation returns the birth time of the node whose birth generates the cavity, which we call the \textit{birth time of the persistent cavity}. The persistent homology computation also returns the birth time of the node at which the cavity becomes tessellated, which we call the \textit{death time of the persistent cavity}. The collection of persistent cavities in dimension $k$ is called the \textit{barcode} in dimension $k$. We can plot the barcode as a series of horizontal lines in which each line corresponds to a persistent cavity and spans from the birth to the death time of the persistent cavity (Fig.~\ref{fig:ph_introduction}E). Each such horizontal line is called a \textit{bar}. Thus, given any undirected, binary growing graph, we are able to find all topological cavities that emerge throughout the growth process and determine the stage of growth at which they emerge and collapse (if they do in fact collapse, otherwise we recover the fact of their indefinite persistence).

\subsection*{Topological Similarity}
The mathematics behind persistent homology allow us to quantitatively investigate the extent to which the persistent homology changes when the birth times of nodes are perturbed. Theoretical work from Ref. \cite{cohen2007stability} gives us an upper-bound for how much the persistent homology can change from this type of perturbation. If the evolving topology of a system changes much less than maximally possible from the birth time reordering, then we say it is a reorderable growth process.

\begin{figure*}[ht]
\centering
\includegraphics[width= 425 px]{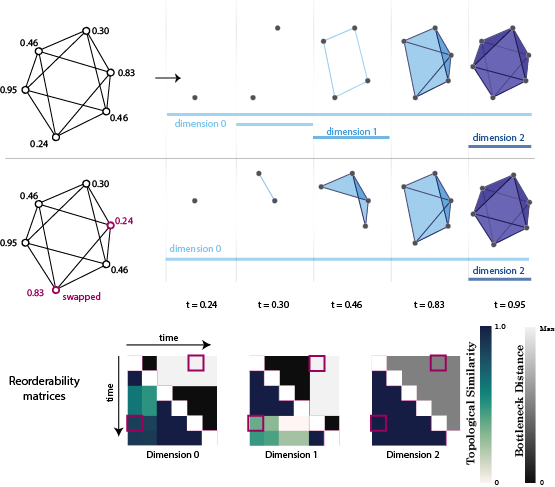}
\caption{\textbf{Calculation of topological similarity.} Here we display the next steps in the analysis of the connectome following on from Fig.~\ref{fig:ph_introduction}. The top row shows the graph G from Fig.~\ref{fig:ph_introduction}, along with its clique complexes and barcode. In the second row, we show the graph that would result from swapping the birth times of the indicated neurons, as well as the corresponding clique complexes and barcode. In the third row, we show the topological similarity and bottleneck distance heatmaps for dimensions zero, one, and two. For any given dimension, the topological similarity between the $i^{th}$ and $j^{th}$ neuron born is found by going down $i$ blocks and right $j$ blocks ($j$ down and $i$ right if $j < i$). The bottleneck distance between these two is found at the mirrored position on the heatmap. The main diagonal is colored white to indicate that both the topological similarity and bottleneck distance of a node with itself is trivial.}
\label{fig:topological_similarity_supplement}
\end{figure*}

More rigorously, for any node pair $v$ and $u$ we can construct a growing graph in the same manner as above, except instead of using the same node-weighted graph $G = (V, E, b)$ we use $G_{v,u} = (V,E,b[_{u}^{v}])$, where $b[_{u}^{v}]$ is identical to $b$ on all values aside from $v$ and $u$, where $b[_{u}^{v}] (v) = b(u)$ and $b[_{u}^{v}](u) = b(v)$ (Fig.~\ref{fig:topological_similarity_supplement}). Using results from Ref. \cite{cohen2007stability}, we know that the distance between this node-swapped growing graph's persistent homology and the persistent homology of $G$ can be no more than $|b(v) - b(u)|$ when we use the bottleneck distance to determine distance between barcodes. Briefly, the bottleneck distance between two barcodes is the maximum difference between birth or death times of matched bar pairs, where we assume a matching of bars that minimizes this difference. For further details, we refer the reader to the Supplementary Methods. 

We denote the bottleneck distance between barcodes in dimension $k$ generated from the original growing graph $G$ and the growing graph of $G_{v,u}$ by $d_{BN}^k(G,G_{v,u})$. By the stability theorem \cite{cohen2007stability}, we have $d_{BN}^k(G,G_{v,u}) \leq |b(v) - b(u)|$. If the $b(v)$, $b(u)$ swap changes the barcode little with respect to the amount it could have changed, then we call $v$ and $u$ topologically similar since they contribute similarly to the persistent homology. In contrast, if the swap changes the barcode to a near maximum amount, then we say $v$ and $u$ are topologically dissimilar. Formally, we calculate the topological similarity between nodes $v$ and $u$ in dimension $k$ as $T_k(v,u) = 1-\frac{d_{BN}^k(G,G_{v,u})}{|b(v) - b(u)|}$ so that node pairs with $T_k(v,u)$ close to 1 are topologically similar and those with $T_k(v,u)$ close to 0 are topologically dissimilar. 

\subsection*{Growing network null models}

Here we utilize four null models for constructing growing graphs: the Erd\H{o}s-R\'{e}nyi growing graph, a spatial growth model, a configuration model, and the economical spatiotemporal growth model \cite{nicosia2013phase}. For the Erd\H{o}s-R\'{e}nyi model, we generate a random growing graph constrained to have the same node number, birth times, and approximate edge density as the true growing connectome. We generate these random graphs by spawning nodes one at a time, and connecting each new node to previously added nodes with probability $\frac{2287}{\binom{279}{2}} \approx 0.59$, which is the edge density of the true \textit{C. elegans} connectome. Nodes are assigned birth times randomly from the list of true neuron birth times. 

To construct the spatial growth model, we first randomly spawn 279 points in $[0,1000] \times [0,40]$, as the adult \textit{C. elegans} is approximately 1000 $\mu$m long with a diameter of 40 $\mu$m \cite{yuan2016terrain}. We then connect the closest 2287 node pairs to match the edge density of the true connectome. Finally, we assign birth times randomly from the list of true neuron birth times. 

To construct the configuration model, we start with the original graph $G$. We note the degree of each node. Then, we remove all the edges. We finally add new edges one at a time between nodes; each edge is placed randomly with the constraint that the final degree of each node is identical to its degree in $G$.

To construct the economical spatiotemporal growth (ESTG) model, we add new nodes $n_i$ one at a time according to their original birth order. At each node addition, we also add connections to each previously spawned neuron $n_j$ with probability $\frac{h_j}{max({h_j | \forall j})} e^{- \frac{d_{i,j}(t)}{0.0126}}$, where $h_j$ is the degree of node $n_j$ in the final connectome, and $d_{i,j}(t)$ is an approximation of the distance between node $n_i$ and $n_j$ at the time when node $n_i$ is added. The approximation is based on reported lengths throughout development, and linear scaling of neuron locations from locations in the adult worm. This model was first defined in Ref. \cite{nicosia2013phase}, where it was shown to reproduce the degree distribution, wiring cost, and node efficiency of the true growing connectome.

\subsection*{Statistical testing}
In the ``Birth-Time Noise'' section we tested the hypothesis that the average distance between the \textit{C. elegans} Betti curves comes from a distribution centered at zero. To test this hypothesis, we used a one-sample $t$-test, since the distances are all independently randomly generated (given the independently random generation of surrogate curves) and ought to come from an approximately normal distribution. The rejection of this hypothesis would indicate that the surrogate curves tend to non-trivially stray from the true curve, and that the Betti curve is meaningfully altered by the introduction of birth-time noise.

In that same section, we tested whether there was a correlation between the Betti numbers of the original curve and the surrogate curves produced by birth-time distortions. To do so, we calculated a Spearman correlation coefficient between the Betti numbers, recorded every minute and in each dimension. The variables were paired ordinal observations with a monotonic relationship, so the Spearman correlation coefficient approach was warranted. The demonstration of such a correlation would indicate that the surrogate curves tend to track the original with strength proportionate to the strength of the correlation.

In the ``Node-Order Noise'' section, we tested whether there was a correlation between the time between two neurons' birth times and their topological similarity. To do so, we calculated a Spearman correlation coefficient between the so-called temporal distance and the topological similarity of all pairs of neurons. The variables were paired ordinal observations with a monotonic relationship, so the Spearman correlation coefficient approach was warranted. The rejection of the null hypothesis would suggest that neurons born temporally far from each other are less topologically reorderable than those born temporally close to each other.

In that same section, we tested the hypothesis that the topological similarities of pairs of neurons born on the same side of the time gap between the $201^{st}$ and the $202^{nd}$ neuron come from the same distribution as the topological similarities from those pairs matched across the divide. To do so, we performed a Mann-Whitney $U$-test on the two sets of topological similarity scores in question. The variables were two sets of independent observations from similar distributions, so the test was warranted. The rejection of this hypothesis would indicate that these sets of pairs of neurons are not equally reorderable, and that one set (neurons matched across the divide) tend to be less reorderable.

In that same section, we tested whether there was a correlation between the time between two neurons' birth times and their topological similarity after the removal of the time gap. To do so, we calculated a Spearman correlation coefficient between the so-called temporal distance and the topological similarity of all pairs of neurons, calculated after modification of their birth times. The variables were paired ordinal observations with a monotonic relationship, so the Spearman correlation coefficient approach was warranted. The demonstration of such a relationship would indicate that neurons born temporally far from each other are less topologically reorderable than those born temporally close to each other, even after the questionably artificial influence of the time gap.

In that same section, we tested the hypothesis that the topological similarities of pairs of neurons born on the same side of the time gap between the $201^{st}$ and the $202^{nd}$ neuron come from the same distribution as the topological similarities from those pairs matched across the divide, even after the influence of the time amount itself had been removed. To do so, we performed a Mann-Whitney $U$-test on the two sets of topological similarity scores in question, calculated after removal of the time gap. The variables were two sets of independent observations from similar distributions, so the test was warranted. The rejection of this hypothesis would indicate that the significance discovered in the previous test was not merely a mathematical artifact of the large time gap, but that there is in fact some topological difference between the neuron pairs in question.

In the ``Comparisons to common growing network null models'' section, we tested the hypothesis that the null models produce connectomes whose Betti curves demonstrate the same resilience to birth-time variation as the true connectome. To do so, we performed a two-sample $t$-test between a set of average distances between null model curves and surrogate curves constructed by permutation of their birth times and a set of average distances between the true curve and surrogate curves constructed by permutation of its birth times by an identical process. The data was two independent sets of continuous variables coming from normally distributed populations, so the test was warranted. The rejection of this hypothesis would demonstrate that the null models produce connectomes with greater (or in fact lesser) resilience to birth-time noise than the true connectome.

In the same section, we tested the hypothesis that the null models produce connectomes where the topological similarity scores for any pair of neurons produces a distribution centered around the true score. To do so, we performed a one-sample $t$-test on a set of differences between null model topological similarity scores and the similarity score exhibited by the pair of neurons in the actual connectome. The data points were independently randomly generated by the growth process, and ought to come from an approximately normal distribution. The rejection of this hypothesis would demonstrate that the null models produce connectomes with topological similarity signatures distinct from the true connectome.

In the ``Relations to Biological Properties'' section, we wished to determine whether there were any significant correlations between a neuron's type/region and its summed topological similarity score. To do so, we performed an analysis of variance (ANOVA) in each dimension to assess relations between the summed similarity and the variables in each biological division (location and type). The summed similarity scores are all nearly independent variables sampled from normal and equivariant populations, so the test was warranted. The results of this ANOVA are useful for determining whether any sort of neuron tends to be more topologically similar overall than other neurons, indicating potential structural importance within the neuron type.

In the ``Sensitivity'' section, we wished to determine whether perturbations of neuron birth-times within a range of reporting errors induced a significantly different Betti curve. To do so, we performed a $t$-test on the set of areas between the base curve and the perturbed curves. We performed this test in each dimension, and for two different sorts of surrogate curve sets. The distances between curves were independently and randomly generated by the surrogate construction process and come from a normal distribution, so the test was warranted. The results of this test would help gauge the impact of the variability with the neuron birth-time reporting.

\section*{Acknowledgements}
We are grateful to Dr. Xiaosong He, Shubhankar Patankar, and Dale Zhou for helpful comments on an earlier version of this manuscript. A.H. and D.S.B. acknowledge the Benjamin Franklin Scholars Program, through which they had the opportunity to meet and initiate this collaboration. A.L., A.S.B., and D.S.B. acknowledge support from the John D. and Catherine T. MacArthur Foundation, the Alfred P. Sloan Foundation, the Institute for Scientific Interchange Foundation, the Paul G. Allen Family Foundation, and the Army Research Office (W911NF-16-1-0474). The content is solely the responsibility of the authors and does not necessarily represent the official views of any of the funding agencies.

\section*{Citation Diversity Statement} 
Recent work in several fields of science has identified a bias in citation practices such that papers from women and other minority scholars are under-cited relative to the number of such papers in the field \cite{mitchell2013gendered,dion2018gendered,caplar2017quantitative, maliniak2013gender, dworkin2020extent,fulvio2021imbalance,chatterjee2021gender,wang2021gendered,bertolero2021racial}. Here we sought to proactively consider choosing references that reflect the diversity of the field in thought, form of contribution, gender, race, ethnicity, and other factors. First, we obtained the predicted gender of the first and last author of each reference by using databases that store the probability of a first name being carried by a woman \cite{dworkin2020extent,zhou_dale_2020_3672110}. By this measure (and excluding self-citations to the first and last authors of our current paper), our references contain 14.65\% woman(first)/woman(last), 15.4\% man/woman, 34.01\% woman/man, and 35.95\% man/man. This method is limited in that a) names, pronouns, and social media profiles used to construct the databases may not, in every case, be indicative of gender identity and b) it cannot account for intersex, non-binary, or transgender people. Second, we obtained predicted racial/ethnic category of the first and last author of each reference by databases that store the probability of a first and last name being carried by an author of color \cite{ambekar2009name, sood2018predicting}. By this measure (and excluding self-citations), our references contain 12.0\% author of color (first)/author of color(last), 16.51\% white author/author of color, 20.88\% author of color/white author, and 50.61\% white author/white author. This method is limited in that a) names and Florida Voter Data to make the predictions may not be indicative of racial/ethnic identity, and b) it cannot account for Indigenous and mixed-race authors, or those who may face differential biases due to the ambiguous racialization or ethnicization of their names.  We look forward to future work that could help us to better understand how to support equitable practices in science.

\clearpage
\newpage
\section*{Supplement}
\setcounter{figure}{0}

\subsection*{Sensitivity analyses}
It is important to perform sensitivity analyses to determine whether errors in the reporting of the latent connectome or in neuron birth times could have impacted our findings and the conclusions we draw therefrom. Indeed, the persistent homology of a growing graph can change if the birth order of nodes is changed and if edges are either added or removed. To determine the robustness of our results with respect to reporting errors, we constructed growing connectomes that took into account (i) the reported failings to locate all connections between neurons and (ii) the margins of error for reported neuron birth times. An estimated 10\% of actual connections are presumed missing due to both incomplete sampling and difficulties of sampling methods \cite{varshney2011structural}. Furthermore, birth times of neurons are also reported with a margin of error of 10\% for neurons born in the embryonic period and 2\% for neurons born afterwards \cite{varier2011neural}. 

\begin{figure*}[ht]
\makeatletter 
\renewcommand{\thefigure}{S\@arabic\c@figure}
\makeatother
\centering
\includegraphics[width=\linewidth]{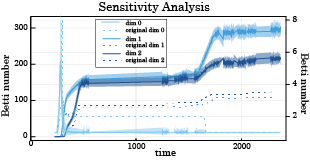}
\caption{\textbf{Robustness of findings to potential errors in connectome construction.} The average of the Betti curves for 1000 surrogate connectomes where birth times were randomly reassigned within margins of reporting error and 800 edges were randomly added. Around each Betti curve is a strip corresponding to $\pm$ the standard deviation across surrogates of the variable being plotted. We further show the Betti curves for the true connectome in dotted lines where in each dimension the same color is used for the null models and the original data.}
\label{fig:reporting_error_one}
\end{figure*}

To implement appropriate sensitivity analyses, we began with the connectome described in the Methods section of the main manuscript. We then constructed 1000 surrogate connectomes by randomly adding 800 connections, some of which were redundant to the graph since multiple connections between a pair of neurons only results in one edge. We also randomly reassigned each neuron's birth time within its original birth time's margin of error. The Betti curves of each such growing connectome were then determined, and averaged into a single set of Betti curves (Fig.~\ref{fig:reporting_error_one}).

The dimension zero curves are nearly identical in the surrogate and original data, aside from the fact that the small second connected component sometimes does not form (Fig.~\ref{fig:reporting_error_one}A). The Betti curves for dimension one and two are qualitatively similar to the true Betti curves. In the embryonic phase, the dimension one curve has an initial spike followed by an interval of lower slope, while the dimension two curve has an initial small jump followed by a much steeper incline. The second dimensional Betti number does not exceed the first, in contrast to the original graph. Across the board, there are more cavities in these dimensions (approximately double for the bulk of the time), but the qualities of the original Betti curves largely remain true in the surrogate cases. While the curves are qualitatively quite similar, the reported curve does lie quite outside the range of surrogate curves. To perform rigorous statistical testing, we used a $t$-test on the set of distances between curves, based on the hypothesis that this distance would be centered at zero, given that the perturbation was trivial. This test rejected that hypothesis with $p$-value less than $0.001$ in all dimensions, with a test statistic in the range of $t=208-316$.

\begin{figure*}[ht]
\makeatletter 
\renewcommand{\thefigure}{S\@arabic\c@figure}
\makeatother
\centering
\includegraphics[width=\linewidth]{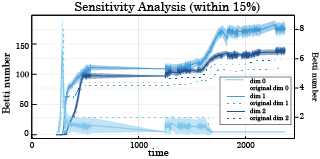}
\caption{\textbf{Robustness of findings to potential errors in connectome construction with restricted wiring cost.} The average of the Betti curves for 1000 surrogate connectomes where birth times were randomly reassigned within margins of reporting error, and 800 edges were randomly added with the restriction that the average connection length remain within 15\% of the original value. Around each Betti curve is a strip corresponding to $\pm$ the standard deviation across surrogates of the variable being plotted. We further show the Betti curves for the true connectome in dotted lines where in each dimension the same color is used for the null models and the original data.}
\label{fig:reporting_error}
\end{figure*}

In constructing these surrogate cases, no attention was paid to the wiring cost of the added connections. Yet, it might be reasonable to expect that the biological nature of the connectome ought to impose some such constraints \cite{rubinov2015wiring}. To account for this biological constraint, we then constructed surrogate connectomes just as before, except we only accepted connectomes whose average wiring cost per connection (calculated as the average connection distance) remained within fifteen percent of the original connectome. The output from 1000 such surrogate graphs is shown in Fig.~\ref{fig:reporting_error}. It appears that the quantitative difference between the original and surrogate graphs largely disappears with this restriction, although again the surrogate Betti curves are distinct from the reported connectome's curve with $p$-values less than $0.001$ and test statistics ranging from $t=45-205$. Collectively, the results of our sensitivity analyses suggest that our original observations will not be significantly undermined by the discovery of the remaining connections or by refinements to neuron birth time reports.

\subsection*{Further methodological details}

\subsubsection*{Persistent homology}
Here we outline the mathematical theory used in calculations for this paper. For a general and more rigorous treatment of persistent homology and algebraic topology in general, we direct the interested reader to Refs. \cite{ghrist2008barcodes,zomorodian2005computing,carlsson2009topology}. 

The intention of using persistent homology is to determine the number of cavities within the structure across various dimensions, and to do so we must construct for each $G_L$ its so-called \textit{clique complex}. For every natural number $n$, a set of $n$ nodes that are all-to-all connected is called an \textit{$n$-clique} Fig.~\ref{fig:cliques_and_cavities}. We may simply think of a clique as a set of nodes for which no subset encloses a cavity. By construction, any subset of nodes within a clique also forms a clique, called a \textit{face} of the clique of which it is a subset. A graph wherein every subset of $n$ all-to-all connected nodes is unified as an $n$-clique is referred to as a \textit{simplicial complex}. More precisely, a simplicial complex $K$ is a set of cliques such that for every $k\in K$, every face of $k$ is in $K$, and the intersection of any pair of cliques is either a face of both cliques or is empty. 

Next, we define a boundary operator $\partial$ on sets of cliques as follows: if $K$ is a set containing one clique, then $\partial K$ is the set of all faces of $K$; if $K$ is a set of multiple cliques, then $\partial K$ is the symmetric difference of all $\partial k$ where $k$ is a one-element subset of $K$ Fig.~\ref{fig:cliques_and_cavities}. A set of (k+1)-cliques whose boundary is empty is called a \textit{k-cycle}. We further define a relation $\sim_k$ on the class of $k$-cycles for each $k$. We say that two $k$-cycles $A$ and $B$ are so related, $A\sim_k B$, whenever their symmetric difference is a boundary of some set of larger cliques Fig.~\ref{fig:cliques_and_cavities}. We wish to show that $\sim_k$ is an equivalence relation. Clearly $A\sim_k A$, and $A\sim_k B$ implies $B\sim_k A$. Suppose $A\sim_k B$ and $B\sim_k C$. Then there exists sets of cliques $K_1$ and $K_2$ such that $\partial K_1 = A \triangle B$ and $\partial K_2 = B \triangle C$. We then have that $\partial ( K_1 \cup K_2) = \partial K_1 \triangle \partial K_2 = (A \triangle B) \triangle (B \triangle C) = A \triangle (B \triangle B) \triangle C = A \triangle \emptyset \triangle C = A \triangle C$. So the symmetric difference of $A$ and $C$ is the boundary of $( K_1 \cup K_2)$ which is a set of higher dimensional cliques, so $A\sim_k C$. Therefore, $\sim_k$ is an equivalence relation on k-cycles. 

The number of non-trivial equivalence classes of the relation $\sim_n$, no member of which is a super-set of a member of another equivalence class, is called the $n^{th}$ \textit{Betti number} ($\beta_n$), or the Betti number of dimension $n$. We can also note that the equivalence classes form a $\mathds{Z}_2$ vector space, and the Betti number of dimension $n$ is the dimension of the vector space of $\sim_n$ equivalence classes; in fact, this is usually the construction used to define the Betti number. Using the notion of Betti number as the dimension of the vector space of equivalence classes, it is simple to show that the $n^{th}$ Betti number is the number of \textit{cavities} of dimension $n$ present in the simplicial complex (Fig.~\ref{fig:cliques_and_cavities}). We will henceforth use the term `equivalent' to mean `related by $\sim_k$ for some k,' and the term `equivalence class' to mean `equivalence class of the relation $\sim_k$.' Note that all Betti numbers $\beta_{n\geq 1}$ are $0$ and $\beta_{0} = 1$ for $G_1$, since by construction this first graph contains a single node and no edges. 

In this way, for every $G_L$ we have a simplicial complex, giving us a sequence of simplicial complexes in which each is a sub-complex of the next. Such a sequence of simplicial complexes is called a \textit{filtered simplicial complex}. For any $L = 1,\dots, N$, every clique present in $G_L$ is also present in $G_{L+1}$ (assuming $1 \leq L \leq 171$). In other words, a clique never disappears from the sequence of complexes once it appears; rather, we merely add more cliques as we move through the sequence. However, equivalence classes of $k$-cycles can both appear and disappear. An equivalence class of $k$-cycles appears in $G_L$ precisely when there exists a $k$-cycle in $G_L$ which is not equivalent to any $k$-cycles in $G_L$ that were present in $G_{L-1}$. An equivalence class of $k$-cycles disappears at $G_L$ precisely when no $k$-cycles within that class in $G_{L-1}$ are present in $G_L$. In other words, such an equivalence class appears whenever a new cavity forms in the simplicial complex, and an equivalence class disappears whenever the addition of a new node cones-off a cavity. 

We can now define a key object of interest for the analyses in this paper: the \textit{barcode} of a filtered simplicial complex. The barcode is a representation of the persistent homology present in a filtered simplicial complex. We first define a \textit{bar} of an equivalence class for some filtered simplicial complex to be an ordered triple of numbers where the first number is the largest label on a node in the simplicial complex wherein the equivalence class is first present; the second number is the smallest label on a node that is not present in the final simplicial complex wherein the equivalence class is present; and the third number is the degree of the cycles in the equivalence class (that is, the third number is the $k$ of the relevant $\sim_k$). In this way, the first number is the birth time of the neuron(s) whose addition gave rise to the cavity; the second number is the birth time of the neuron(s) whose addition gave rise to the termination of the cavity; and the third number is the dimension of the cavity. If the cavity persists indefinitely, we fill the second index with $\infty$. The \textit{barcode} of a filtered simplicial complex is simply the set of all its bars, thus containing all the information relevant to the complex's persistent homology. The barcode is often given a graphical representation where for each bar there is a line spanning from the first number of the bar to the second number. Each bar is then colored to indicate the dimension of the cycles in its equivalence class. In fact, barcodes tend to be defined as this graphical representation, which stores information equivalent to the construction used in this paper.

\subsubsection*{Reorderabiltiy}
The reorderability of a filtered simplicial complex is calculated by analysing some metric of reorderability for each pair of nodes. The calculation used here is identical for any pair, so we will describe how the reorderability measure is calculated for the node pair $(a,b)$. By construction, node $a$ and node $b$ both have labels equal to the birth times of their corresponding neurons. We define the graph $G_{a,b}$ to be identical to graph $G$ except the labels on node $a$ and node $b$ are swapped, such that node $b$ bears the birth time of node $a$'s neuron, and \emph{vice versa}. We then construct a telescoping sequence of graphs in a manner identical to the manner used above except using $G_{a,b}$ instead of $G$. In the same way, we construct the filtered simplicial complex and Betti numbers for $G_{a,b}$. Thus, we have the barcodes for the filtered simplicial complex that would result if the neurons corresponding to node $a$ and $b$ had been born at each other's birth times. Note that if $a$ and $b$ have the same birth time, then the resulting barcodes will be identical to those of the original complex, since the $G_{a,b}$ and $G$ will be identical.

We define a distance between barcodes as follows: Let $B_1$ and $B_2$ be two barcodes, and let $\gamma$ be a bijection between a pair of sets $B_1'$ and $B_2'$, which are the same as $B_1$ and $B_2$ (respectively) except that they only contain bars of dimension $n$, and each contain infinitely many copies of $\{ (r,r) | \forall r \in \mathds{R} \}$. Thus, each bar in each barcode is paired with either a bar in the other barcode, or with a pair of identical real numbers. The inclusion of the $(r,r)$ pairs should be seen as including infinitely many bars of length zero. We define the \textit{bottleneck distance} between two barcodes, $B_1$ and $B_2$, in dimension $n$ as follows: $d_{BN}^n(B_1, B_2) = min_{\gamma}( max_{b \in B_1'} ( \{ |b_1 - \gamma(b)_1|, |b_2 - \gamma(b)_2|\} ) )$, where $b_k$ is the $k$th of the three numbers that make up the bar $b$. In other words, we first define the distance between two bars to be either the difference in their initial points or the difference in their terminal points, whichever is larger. We then imagine all possible pairings of bars between two barcodes, where bars are allowed to be paired with the \textit{trivial bars} of the other barcode; trivial bars are those with identical initial and terminal points. The distance between the two barcodes is then the greatest distance between bars in whichever pairings produces the smallest such greatest distance. We may interpret the selection of $\gamma$ as choosing the best pairing of bars, given that we wish to minimize the greatest distance observed between all paired bars' initial and terminal values. 

Now we wish to show that $d_{BN}^n$ is a metric to warrant its description as a distance. Clearly $d_{BN}^n(B_1, B_1) = 0$, as can be seen by selecting the trivial $\gamma$. Suppose $d_{BN}^n(B_1, B_2) = d$, meaning over all bijections there is a minimizing $\gamma$ such that $max_{b \in B_1'} ( \{ |b_1 - \gamma(b)_1|, |b_2 - \gamma(b)_2|\} = d$. But then $$max_{b \in B_2'} ( \{ |b_1 - \gamma ^{-1}(b)_1|, |b_2 - \gamma^{-1}(b)_2|\}$$  $$= max_{b \in B_1'} ( \{ |\gamma ^(b)_1 - \gamma ^{-1}(\gamma (b))_1|, |\gamma (b)_2 - \gamma ^{-1}(\gamma(b))_2|\} $$  $$= max_{b \in B_1'} ( \{ |\gamma ^(b)_1 - b_1|, |\gamma (b)_2 - b_2|\} $$  $$= max_{b \in B_1'} ( \{ |b_1 - \gamma(b)_1|, |b_2 - \gamma(b)_2|\} = d$$ Suppose there was a bijection $\Gamma$ such that $$max_{b \in B_2'} ( \{ |b_1 - \Gamma(b)_1|, |b_2 - \Gamma(b)_2|\} < max_{b \in B_1'} ( \{ |b_1 - \gamma^{-1}(b)_1|, |b_2 - \gamma^{-1}(b)_2|\}$$ Then, by a similar argument as above, we would have that $$max_{b \in B_1'} ( \{ |b_1 - \Gamma^{-1}(b)_1|, |b_2 - \Gamma^{-1}(b)_2|\} < max_{b \in B_1'} ( \{ |b_1 - \gamma(b)_1|, |b_2 - \gamma(b)_2|\}$$ which cannot be the case since by construction $\gamma$ was the bijection producing the minimum possible value. So no such $\Gamma$ exists, and thus $d_{BN}^n(B_2, B_1) = d = d_{BN}^n(B_1, B_2)$. Finally, let $\gamma_1$ and $\gamma_2$ be the minimizing pairings used for $d_{BN}^n(B_1, B_2)$ and $d_{BN}^n(B_2, B_3)$, respectively. We then have that $\Gamma = \gamma_2 \circ \gamma_1$ is a pairing of elements from $B_1$ onto $B_3$. By the triangle inequality of the supremum metric, we have that $$\forall b \in B_1', (max ( \{ |b_1 - \gamma_1(b)_1|, |b_2 - \gamma_1(b)_2|\}) + max ( \{ |\gamma_1(b)_1 - \gamma_2(\gamma_1(b))_1|, |\gamma_1(b)_2 - \gamma_2(\gamma_1(b))_2|\})$$  $$ \geq max ( \{ |b_1 - \gamma_2(\gamma_1(b))_1|, |b_2 - \gamma_2(\gamma_1(b))_2|\})$$ We then have, $$d_{BN}^n(B_1, B_2) + d_{BN}^n(B_2, B_3) $$  $$= min_{\gamma}( max_{b \in B_1'} ( \{ |b_1 - \gamma(b)_1|, |b_2 - \gamma(b)_2|\} ) ) + min_{\gamma}( max_{b \in B_2'} ( \{ |b_1 - \gamma(b)_1|, |b_2 - \gamma(b)_2|\} ) )$$  $$ = max_{b \in B_1'} ( \{ |b_1 - \gamma_1(b)_1|, |b_2 - \gamma_1(b)_2|\} ) + max_{b \in B_2'} ( \{ |b_1 - \gamma_2(b)_1|, |b_2 - \gamma_2(b)_2|\} ) $$  $$=  max_{b \in B_1'}( \{ |b_1 - \gamma_1(b)_1|, |b_2 - \gamma_1(b)_2|\}) + max_{b \in B_1'}( \{ |\gamma_1(b)_1 - \gamma_2(\gamma_1(b))_1|, |\gamma_1(b)_2 - \gamma_2(\gamma_1(b))_2|\}) $$  $$\geq max_{b \in B_1'}(max ( \{ |b_1 - \Gamma(b))_1|, |b_2 - \Gamma(b))_2|\}) $$  $$\geq d_{BN}^n(B_1, B_3)$$ We thus have shown that $d_{BN}^n$ is symmetric, reflexive, and satisfies the triangle inequality, so it is therefore a metric and may properly be called a distance function. 

With this bottleneck distance in dimension $n$ defined for pairs of barcodes, we define the bottleneck distance in dimension $n$ for pairs of nodes to be the bottleneck distance in dimension $n$ between the barcodes of the original filtered simplicial complex and the barcodes of the filtered simplicial complex generated by swapping the labels on the nodes; that is, that the bottleneck distance between nodes $a$ and $b$ is $d_{BN}^n(G,G_{a,b})$. We may refer to the bottleneck distance between a pair of nodes without referring to a dimension, in which case we mean the averaged bottleneck distance over dimensions zero through two. Note that the bottleneck distance between two neurons with identical birth times will be zero.

We define another measure based on this bottleneck distance called the \textit{topological similarity}. The topological similarity in dimension $n$ of a pair of nodes $a$ and $b$ with labels $t_1$ and $t_2$, respectively, is defined as $T_n(a,b) = 1 - \frac{d_{BN}^n(a,b)}{|t_1 - t_2|} $. By the bottleneck stability theorem, $d_{BN}^n(a,b) \leq |t_1 - t_2|$, so $T_n(a,b)$ is always within $[0,1]$, with low bottleneck distance resulting in high topological similarity, and \emph{vice versa} \cite{cohen2007stability}. We may refer to the topological similarity between a pair of nodes without referring to a dimension, in which case we mean the averaged topological similarity over dimensions zero through two. Note also that the topological similarity between two neurons with identical birth times will be 1. 

\subsubsection*{Imposing Linear Order on Birth Times}
At one point in our analysis, it was necessary to determine the number of cavities born or killed by each neuron. Given that 107 out of the 279 neurons have non-unique birth times, the above method of constructing a growing graph would preclude this sort of analysis. In order to use the software Eirene to measure the effect that each neuron has individually, a linear ordering of neuron birth times is needed. We define an \textit{imposed random linear ordering on birth times} as follows: for any time $t$ at which $n>1$ neurons are born, we randomly assign each such neuron an integer in $k \in [1,n]$, and modify its birth time by adding $\frac{k}{n*100+1}$ to its original birth time. Since the minimum distance between two non-identical birth times is no greater than $0.01$ (due to the fact that birth times are reported to two decimal places), and since $\frac{k}{n*100+1} \leq \frac{n}{n*100+1} < \frac{1}{100+1} < \frac{1}{100}$, we have that neurons originally born at distinct times maintain their relative ordering. 

This method of imposing a linear order allows us to determine the number of cavities born or killed by each neuron. Yet, it can also produce artifacts whose effects we must mitigate. For example, the process may generate cavities not present in the original growing graph since adding nodes with identical birth times one at a time may briefly generate cavities that would not be present normally. To address this potential artifact, we ignore such cavities in our calculations. Furthermore, if multiple neurons born at the same time are all necessary for the generation of a cavity, only the node ordered last by the random ordering would be seen as generating the cavity. To try to mitigate the effects of this issue, we performed 1000 trials of random orderings for the relevant calculations. Thus, if $n$ neurons are all necessary for the production of a cavity, each will receive credit for generating the cavity in approximately $\frac{1000}{n}$ times. For each neuron, we determine the standard deviation across the set of cavities it generates (or destroys) in each dimension. We find that, with forming dimension 1 cavities as an exception, the vast majority of neurons generate (kill) an identical number of cavities in each trial, indicating that such neurons always independently create (destroy) the cavities for which they are given credit.


\begin{figure*}[ht]
\makeatletter 
\renewcommand{\thefigure}{S\@arabic\c@figure}
\makeatother
\centering
\includegraphics[width=\linewidth]{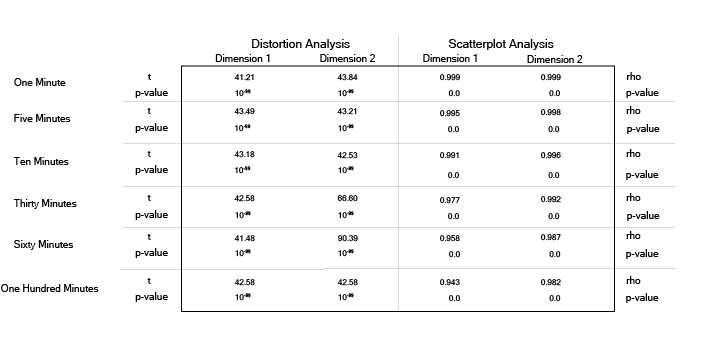}
\caption{\textbf{T-test and Spearman correlation results from analyses presented in Figure 2.} We have displayed the outputs of statistical tests performed on the neuron birth time distorted surrogate curves. On the left, we have the $p$-values for $t$-test of the hypothesis that the signed distances between the true Betti curves and the set of surrogate curves (of the indicated maximum time variation) is a normal distribution centered at 0. Below the $p$-value is a confidence interval for the mean of that average difference. These tests included 1000 surrogate curves for each grouping. On the right we have the Spearman $\rho$ value for the correlation between the true Betti numbers and the surrogate Betti numbers recorded on a minute-by-minute basis.}
\label{fig:noise_birth_times_stats}
\end{figure*}

\begin{figure*}[ht]
\makeatletter 
\renewcommand{\thefigure}{S\@arabic\c@figure}
\makeatother
\centering
\includegraphics[width=\linewidth]{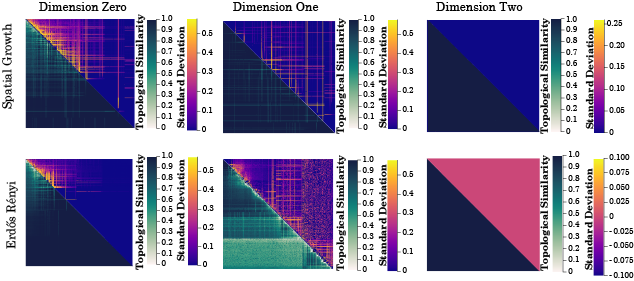}
\caption{\textbf{Persistent homology for the spatial growth and Erd\H{o}s-R\'{e}nyi null models.} For both the spatial growth and the Erd\H{o}s-R\'{e}nyi null models, we have reported the average of the topological similarity across the four surrogate null model outputs (in shades of green) and the standard deviation of the topological similarity value from across the null model outputs (in a map from yellow to purple).}
\label{fig:null_model_top_sim_supp}
\end{figure*}

\begin{figure*}[ht]
\makeatletter 
\renewcommand{\thefigure}{S\@arabic\c@figure}
\makeatother
\centering
\includegraphics[width=\linewidth]{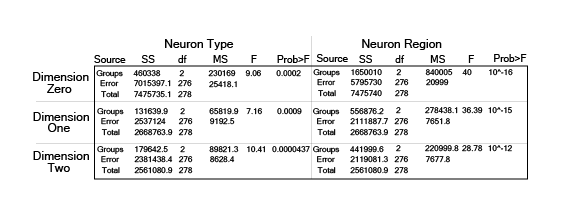}
\caption{\textbf{Statistical results from analyses in Figure 6.} Here we provide the outputs of an ANOVA assessing the relationship between summed topological similarity and biological features of neurons.}
\label{fig:linear_order_trials_supplement}
\end{figure*}

\clearpage
\newpage
\bibliographystyle{vancouver}
\bibliography{bibfile.bib}

\end{document}